\documentclass[pra,showpacs,showkeys,superscriptaddress,nofootinbib]{revtex4}

\usepackage{amsmath,amssymb}
\usepackage{epsfig}
\usepackage{color}
\newcommand{\n}{\nonumber}
\newcommand{\be}{\begin{equation}}
\newcommand{\ee}{\end{equation}}
\newcommand{\bea}{\begin{eqnarray}}
\newcommand{\eea}{\end{eqnarray}}

\begin{document}

\title{\textbf{Similarity solutions of Reaction-Diffusion equation \\
with space- and time-dependent diffusion and reaction terms}}
\author{C.-L. Ho}%\\
%\address
\affiliation{Department of Physics, Tamkang University
Tamsui 25137, Taiwan}

\author{C.-C. Lee}
\affiliation{Center of General Education, Aletheia University, Tamsui 25103, Taiwan}

\date{Aug 1, 2015}

%\maketitle  % for LaTex

\begin{abstract}

We consider solvability of the generalized reaction-diffusion
equation with both space- and time-dependent diffusion and reaction terms
by means of the similarity method. By  introducing  the
similarity variable, the reaction-diffusion equation is reduced to an
ordinary differential equation. Matching the
resulting ordinary differential equation with known exactly solvable equations, one can obtain corresponding exactly solvable reaction-diffusion systems.   Several representative examples of exactly solvable reaction-diffusion equations are presented.
\end{abstract}

% 05.10.Gg  Stochastic analysis methods (Fokker-Planck, Langevin, etc.)
%05.90.+m   Other topics in statistical physics, thermodynamics, and nonlinear dynamical systems (restricted to new topics in section 05)
%02.50.Ey   Stochastic processes


\maketitle

%\newpage

%%%%%%%%%%%%%%%%%%%%%%
\section{Introduction}

Many natural phenomena involve the change of concentration/population of one or more substances/species distributed in space  under the influence of two processes: local reaction which modify the concentration/population, and diffusion which causes the substances/species  to spread in space.   Such phenomena are well modelled by the reaction-diffusion equation (RDE).

The general form of RDE of the concentration $W(x,t)$ of a single component in one spatial dimension is
\be
\frac{\partial }{\partial t}W(x,t)=D\frac{\partial^2}{\partial x^2} W(x,t) + f(W),
\label{KPP}
\ee
where $D$ is the constant diffusion coefficient and $f(W)$ is the reaction term which accounts for the  local reaction.
Eq.~(\ref{KPP}) is also called the KPP (Kolmogorov-Petrovsky-Piscunov) equation, named after  the authors who first studied some of the mathematical properties of the RDE.  This equation encompasses  the diffusion (heat) equation ($f=0$) and the Fokker-Planck equation (when $f$ is a gradient term of some function linear in $W$) \cite{FPE}.

Different forms of the reaction term $f$ have been proposed to describe different phenomena. For instance, the choice $f=W(1-W)$ yields the Fisher equation employed in the study of  wave propagation of advantageous genes in a population \cite{Fisher} and evolution of a neutron population in a nuclear reactor \cite{neutron}. Rayleigh-Bernard convection is studied using RDE with $f=W(1-W^2)$ \cite{fluid}, while combustion and shock waves phenomena invoke RDE with $f=W(1-W)(W-a) (0<a<1)$ \cite{shock}.    Generalization and extension of the KPP equation to higher dimensions and multi-component cases also find  interesting applications in chemical kinetics \cite{Arnold}, pattern formation and morphogenesis \cite{Turing}, nerve pulse propagation in nerve systems \cite{nerve}, and other biological systems \cite{Murray}.

In view of its broad applicability, it is thus desirable to obtain analytic solutions of the RDE for as many systems as possible. However,
just as any equation in sciences, solving the RDE exactly is in general a formidable task, except in a few simplified cases. Fortunately, for many cases of the RDE mentioned before, exact solutions can be found in the form of travelling wave solutions \cite{GK}.

In this paper we would like to consider exact solvability of the RDE in terms of the similarity solutions \cite{BC}. This is motivated by our recent works on
similarity solutions of the Fokker-Planck equation, which as mentioned before is a subclass of the  RDE \cite{Ho1,Ho2,Ho3}.
We found  it very interesting that for a class of the Fokker-Planck equation with time- and space-dependent coefficients,  a general formula of exact solutions can be obtained in closed form by the similarity method, both for fixed and moving boundaries.
One advantage of the similarity method is that it allows one to
reduce the partial differential equation under consideration to an
ordinary differential equation which is generally easier to solve, provided that the original equation
possesses proper scaling property under certain scaling
transformation of the basic variables.   Here we would like to extend our previous consideration to the RDE. It turns out that, similar to the Fokker-Planck case, one can determine the possible functional forms of the diffusion and the reaction term of the RDE in order to get exactly solvable system with similarity solutions.

This paper is organized as follows.
 Sect.~II discusses the
scaling properties of the RDE.
Sect.~III introduces the corresponding similarity variable and scaling forms of the relevant functions, which are used to
 reduce the RDE into an ordinary differential
equation.  The equation of continuity is discussed in Sect.~IV which helps to identify two types of  scaling behaviours of the RDE.
Some examples of these two types of scaling reaction-diffusion (RD) systems are presented in Sect.~V and VI, respectively.
Sect.~VII concludes the paper.

%%%%%%%%%%%%%%%%%%%
\section{Scaling of Reaction-Diffusion equation}

We shall consider the following general form of the RDE in $(1+1)$-dimension
\be
\frac{\partial W(x,t)}{\partial t}=\frac{\partial}{\partial x}\left(D(W,x,t)\frac{\partial}{\partial x} W(x,t)\right) + f(W,x,t),
\label{RDE}
\ee
where $W(x,t)$ is the particle density function,
$D(W,x,t) $ is the diffusion coefficient and $f(W, x,t)$ the
reaction term. We use the term  ``particle" to denote generally the number of basic member of a substance or a specie.  The domains we shall consider in this paper are the real line
$x\in (-\infty,\infty)$, or  the half lines $x\in [0,\infty)$ and
$(-\infty,0]$.  Cases with finite domains, which correspond to systems with moving boundaries, can be considered similarly \cite{Ho2}.  To cater for the most general situation, we leave the possibility that $D$ and $f$ could be functions of $W$.

We shall consider the similarity solutions of the RDE.  Such
solutions are possible, provided the RDE possesses certain
scaling symmetry. Below we shall study the scaling property of the RDE.

Consider the scale transformation
\be
x=\epsilon^a \bar{x}\;\;\;,\;\;\; t=\epsilon^b \bar{t},
\ee
where the scale factor $\epsilon$ and the two scaling exponents $a$ and $b$ are real parameters. Suppose
under this transformation, the density function, the diffusion coefficient and
the reaction term  scale as
\be
W(x,t)=\epsilon^c \bar{W}(\bar{x},\bar{t}),~~
D(W,x,t) =\epsilon^d \bar{D}(\bar{W}, \bar{x},\bar{t}),\;\;\;\;\;\;
f(W,x,t)=\epsilon^e \bar{f}(\bar{W}, \bar{x},\bar{t}).
\label{scale}
\ee
Here the scaling exponents $c$, $d$ and $e$ are also some real parameters.  Written in
the transformed variables, Eq.(\ref{RDE}) becomes
\be\
\epsilon^{c-b}\frac{\partial \bar{W}}{\partial
\bar{t}}=\epsilon^{-2a+c+d}\frac{\partial}{\partial
\bar{x}}\left(\bar{D}\frac{\partial}{\partial \bar{x}}\bar{W}\right) + \epsilon^e\, \bar{f}.
\label{RDE2}
\ee
For simplicity and clarity of presentation, here and below we shall often omit the independent variables in a function.

One sees that if the scaling indices satisfy $b=2a-d=c-e$, then
Eq.(\ref{RDE2}) has the same functional form as Eq.(\ref{RDE}).
In this case, the RDE admits similarity solutions.   We shall
present such solutions below.

%%%%%%%%%%%%%%%%%%%%%%%%%%%
\section{Similarity variable and scaling forms}

The similarity method is a very useful method for solving a
partial differential equation which possesses proper scaling
behavior. One advantage of the similarity method is to reduce the
order of a partial differential equation through some new
independent variables (called similarity variables), which are
certain combinations of the old independent variables such that
they are scaling invariant, i.e., no appearance of parameter
$\epsilon$, as a scaling transformation is performed.

In our case, the second order RDE can  be transformed into an
ordinary differential equation which is generally easier to solve. Here
there is only one similarity variable $z$, which can be defined as
\be
z\equiv\frac{x}{t^{\alpha}}, ~~\mbox{where}~
\alpha=\frac{a}{b}\;\;\;,\; b\neq 0\;.
\ee

Next we assume the following scaling forms of the density function, the diffusion and the reaction terms in terms of $z$:
\be
W(x,t)=t^\mu y(z), ~~ D(W,x,t)=t^\nu \rho(z), ~~ f(W,x,t)=t^\lambda \sigma(z).
\ee
From Eq. (\ref{scale}) together with the scaling conditions $b=2a-d=c-e$,  one has
\be
\mu=\frac{c}{b}, ~~ \nu=\frac{d}{b}=2\alpha-1, ~~\lambda=\frac{e}{b}=\mu-1.
\ee
Thus $\alpha$ and $\mu$ are the only two independent scaling exponents of the RDE.

In terms of these scaling forms, Eq.~(\ref{RDE}) reduces to an ordinary differential equation
\be
\frac{d}{dz}\left(\rho\frac{d}{dz}y\right) + \alpha z\frac{dy}{dy}-\mu y+ \sigma(z)=0.
\label{ODE}
\ee
Note that when $\mu=-\alpha$, which we will encounter below, Eq.\,(\ref{ODE}) reduces to
\be
\frac{d}{dz}\left(\rho\frac{d}{dz}y + \alpha z y\right)+ \sigma(z)=0.
\label{ODE_e}
\ee

To proceed further, we shall consider the conditions imposed by
the continuity in the change of the particle number of the system,
i.e., the equation of continuity.

%%%%%%%%%%%%%%%%%
\section{Equation of continuity}

The total number $N$ of the system is related to the density function $W(x,t)$ by
\be
N=\int_{\cal D}\,W(x,t)\,dx=t^{\alpha+\mu}\int_{\cal D}\,
y(z)\,dz,
\label{N}
\ee
where $\cal D$ is the domain of the independent variable.  For simplicity, we use the same notation $\cal{D}$ for both the variable $x$, and the corresponding similarity variable $z$. 

 Eq.\,(\ref{N}) distinguishes two different situations: $\alpha+\mu\neq 0$ and $\alpha+\mu=0$.
It is obvious from this equation that $N$ is conserved if and only if  $\mu=-\alpha$.

Eq.\,(\ref{N}) implies that
\be
\frac{dN}{dt}=(\alpha+\mu)t^{\alpha+\mu-1} \left(\int_{\cal D}\,
y(z)\,dz\right).
\label{dN}
\ee
On the other hand, from Eq.\,(\ref{RDE}) one has
\bea	
\frac{dN}{dt}&=&\int_{\cal D}\,\frac{\partial W}{\partial t}\,dx
=\Delta\left(D(W,x,t)\frac{\partial}{\partial x} W(x,t)\right)_{\partial {\cal D}}+\int_{\cal {\cal D}}\, f(W,x,t)\,dx\n\\
&=&t^{\alpha+\mu-1} \Delta \left(\rho\frac{dy}{dz}\right)_{\partial {\cal D}}+ t^{\alpha+\lambda}  \int_{\cal D} \sigma(z)\,dz.
\eea
Here $\partial{\cal D}$ denotes the boundaries of the domain $\cal D$, and $\Delta(\cdots)_{\partial {\cal D}}$ the difference of the terms in the bracket at the boundaries.

In view of $\lambda=\mu-1$, one has
\be
 (\alpha+\mu)\int_{\cal D}\,
y(z)\,dz= \int_{\cal D}\, \sigma(z)\,dz + \Delta \left(\rho\frac{dy}{dz}\right)_{\partial {\cal D}}.
\label{EOCa}
 \ee
 If $y(z)\to 0$ fast enough so that the boundary terms tend to zero, then one has
 \be
 (\alpha+\mu)\int_{\cal D}\,
y(z)\,dz= \int_{\cal D}\, \sigma(z)\,dz.
\label{EOC}
 \ee
 This is the situation we shall consider in most of the cases below, except the example in Sect.\,VI.D.

In what follows, we shall present some examples for $\mu=-\alpha$ and $\mu\neq -\alpha$.

%%%%%%%%%%%%%%%%%%%%%
\section{Cases with $\mu= -\alpha$}

 As mentioned at the beginning of Sect.\,IV, $N$ is conserved when  $\mu=-\alpha$.  Hence one can normalize $W(x,t)=t^{-\alpha}y(z)$  and consider it as the probability distribution function.
 
 Furthermore, Eq.~(\ref{EOC}) implies $\int_{\cal D}\, \sigma (z) dz=0$.   This is most easily satisfied if $\sigma(z)$ is a total differential, i.e., $\sigma(z)=-d\tau(z)/dz$ for some function $\tau(z)$. This includes as subclass any function $\sigma(z)$ that is anti-symmetric w.r.t. the mid-point of the domain $\cal{D}$ in the similarity variable $z$.   But this latter situation  is possible only if  $\cal{D}$  is the whole line or a finite domains in the $z$-space  (corresponding to moving boundaries in the $x$-space), and is not possible for the half-line. 
  
 Consequently,  Eq.~(\ref{ODE_e}) is a total derivative, and can be integrated once to give
 \be
\rho y'+ \alpha z y-\tau={\rm constant}.
\label{ODE_c}
\ee
 Here the prime denotes derivative w.r.t $z$.
 For $y, \tau\to 0$ at the boundaries,  the constant equals zero and we need only to consider the following equation instead
  \be
\rho y'+ \alpha z y-\tau=0.
\label{ODE1}
\ee

% ------    FK type --------
\subsection{Fokker-Planck type}

First we consider the situation in which the function $\tau(z)$  is proportional to $y(z)$, i.e.
$\tau(z)=\beta(z)y(z)$ for some function $\beta(z)$. In this case the RDE is of the Fokker-Planck type, where the function $\beta(z)$ plays the role of the drift coefficient.   Integrating Eq.~(\ref{ODE1}) once gives
\be
y(z)\propto \exp\left(\int^z\,dz \, \frac{\beta(z)-\alpha z}{\rho(z)}\right).
\label{FK}
\ee
Hence for any choice of $\beta(z)$ and $\rho(z)$ such that $y(z)$ in Eq.\,(\ref{FK}) is integrable and that $W(x,t)$ is normalizable, one has an exactly solvable RD system. This is exactly the same as the way to obtain similarity solutions of the Fokker-Planck equations discussed in \cite{Ho1,Ho2,Ho3}.   All the cases presented there for the Fokker-Planck equations  can be carried over to this type of RDE.
As such we shall be brief on this case, and present only an   example for illustration.

%--------
%\subsubsection{Diffusing wave with moving peak}

Let us take $\rho(z)=1$ and $\beta(z)=\beta_1 z+\beta_0$ ($\beta_1<\alpha$).  Then
the system is given by
$D(x,t)=t^{2\alpha-1},~f(x,t)=-t^{-(\alpha+1)}d\tau(z)/dz$, and
\be
W(x,t)=\sqrt{\frac{\alpha-\beta_1}{2\pi t^{2\alpha}}}\,
e^{-\frac12\frac{\alpha-\beta_1}{t^{2\alpha}}
\left(x-\frac{\beta_0}{\alpha-\beta_1}t^\alpha\right)^2},~~\alpha>\beta_1,
\ee
for $t\geq 0$ and $-\infty<x<\infty$.
This solution represents a diffusing wave wave with  a moving peak.

Other possible forms of $y(z)$, including those related to solutions with moving boundaries, and solutions involving the recently discovered exceptional orthogonal polynomials, can be found in Ref.\,\cite {Ho2} and \cite{Ho3}, respectively.

%-------     Non FK type --------
\subsection{Non-Fokker-Planck type}

We now consider the situation where $\sigma(z)$ is not proportional to $y(z)$, i.e., $\sigma(z)\neq y(z)$.

Again we assume $y, \tau\to 0$ at the boundaries,  so the constant in Eq.\,(\ref{ODE_c}) equals zero and we need consider only Eq.\,(\ref{ODE1}).
The general solution of Eq.\,(\ref{ODE1}) is
\be y(z)=e^{-\alpha\int^z
\frac{z}{\rho}dz}\left(\int^z e^{\alpha\int^z
\frac{z}{\rho}dz}\frac{\tau}{\rho}dz +C\right), ~~C={\rm
constant}.
\label{GS}
\ee
 Any choice of $\rho(z)$ and $\tau(z)$ such that $y(z)$ is exactly integrable  and $W(x,t)$ is nomalizable furnishes a solvable RD system.

Some examples below serve to elucidate the idea.

 %--------------
 \subsubsection{$\rho(z)=1$}

%\be
%y(z)=e^{-\frac{\alpha}{2}z^2} \left(\int^z e^{\frac{\alpha}{2}z^2}\tau(z)dz+ C\right).
%\ee

%Take $\tau(z)=2\gamma z \exp(-\tau z^2)$ with real constants $\gamma, \tau>0$.  The solution is

Take $\tau(z)=2\gamma z \exp(-\eta z^2)$ with real constants
$\gamma, \eta>0$. The general solution $y(z)$ in Eq.(\ref{GS}) is
\be
 y(z) = \frac{2\gamma}{\alpha -2\eta}e^{-\eta
 z^2}+Ce^{-\frac{\alpha}{2}z^2},
\ee for $z\in(-\infty,\infty)$. Thus we obtain an exactly solvable
RD system with \bea
 D(x,t) &=& t^{2\alpha -1}, \n\\
 f(x,t) &=& -2\gamma \left( 1-2\eta \left(\frac{x}{t^\alpha}\right)^2\right)
 t^{-(\alpha+1)}e^{-\eta(\frac{x}{t^{\alpha}})^2}, \\
 W(x,t) &=& t^{-\alpha}\left( \frac{2\gamma}{\alpha -2\eta}e^{-\eta
 (\frac{x}{t^{\alpha}})^2}+Ce^{-\frac{\alpha}{2}(\frac{x}{t^{\alpha}})^2}
 \right).\n
\eea
To ensure $W(x,t)\geq 0$ for all $x\in (-\infty, \infty)$, we must have $C\geq \mp2\gamma/|\alpha-2\eta|$ for $\alpha>2\eta$ and $\alpha<2\eta$, respectively.

In Fig.1 we show the graphs of $D(x,t)$, $f(x,t)$ and $W(x,t)$ for
a set of parameters with three different times.

%---------------

\subsubsection{$\rho(z)=z$}

Let $\tau(z)=z \exp(-\eta z)$ with real constants $\eta>0$. The
general solution $y(z)$ is 
\be
 y(z) = \frac{1}{\alpha -\eta}e^{-\eta z}+Ce^{-\alpha z},
\ee for $z\in [0,\infty)$.

The corresponding exactly solvable RD
system is defiined by
\bea
 D(x,t) &=& t^{\alpha -1}x, \n\\
 f(x,t) &=& -\left( 1-\eta \left(\frac{x}{t^{\alpha}}\right) \right)
 t^{-(\alpha +1)}e^{-\eta (\frac{x}{t^{\alpha}})},\\
  W(x,t) &=& t^{-\alpha}\left( \frac{1}{\alpha -\eta}e^{-\eta (\frac{x}{t^{\alpha}})}+Ce^{-\alpha (\frac{x}{t^{\alpha}})}
 \right).\n
\eea
To ensure $W(x,t)\geq 0$ for all $x\in [0,\infty)$, we must have $C\geq \mp1/|\alpha-\eta|$ for $\alpha>\eta$ and $\alpha<\eta$, respectively.

We show in Fig.\,2 the graphs of $D(x,t)$, $f(x,t)$ and $W(x,t)$ for
a set of parameters with three different times.

%----------------
\subsubsection{$\rho(z)=\alpha z^2$}

For $\rho(z)=\alpha z^2$, Eq.\,(\ref{GS}) becomes
\be
y(z)=\frac{1}{z}\left(\int^z \frac{\tau(z)}{\alpha z}dz +C\right).
\ee
One must set $C=0$ in order that $y(z)$ is finite at $z=0$.

As an example, let us take $\tau(z)$ such that $y(z)=\exp(-\beta z^2), \beta>0$.
This means
\be
\tau(z)=\alpha z \left(1-2\beta z^2\right)e^{-\beta z^2}.
\ee
The corresponding RD system is defined by
 \bea
 D(x,t) &=& \frac{1}{2}\alpha \ t^{-1} x^2,~~0\leq x <\infty, \n\\
  f(x,t) &=& \alpha t^{-(\alpha+1)} \left[6\beta \left(\frac{x}{t^\alpha}\right)^2 + 2\beta \left(\frac{x}{t^\alpha}\right) -1\right] e^{-\beta (\frac{x}{t^\alpha})^2},\\
 W(x,t) &=& t^{-\alpha} e^{-\beta (\frac{x}{t^\alpha})^2}.\n
\eea

It is interesting to see that in a rather complicated reaction term $f(x,t)$ leads to a simple Gaussian-like distribution $W(x,t)$.

%----------   Non linear diffusion -----------------------------------
\subsection{A nonlinear diffusion equation}

Taking $\tau(z)=0$ and $\rho(z)=y^n(z)$ ($n=0,1,2,\ldots$) , one gets a nonlinear equation
\be
\frac{d}{dz}\left(y^n(z)\frac{d}{dz}y(z)+\alpha z y(z)\right)=0.
\label{I-3}
\ee
This is the reduced scaling ODE associated with the nonlinear diffusion equation
\be
\frac{\partial W(x,t)}{\partial t}=\frac{\partial}{\partial x}\left(W^n(x,t)\frac{\partial}{\partial x} W(x,t)\right),
\label{NDE}
\ee
where the diffusion coefficient is given by $D(x,t)=W^n(x,t)$ \cite{Munier}.

In \cite{Munier},  Eq.\,(\ref{NDE}) was solved through a very indirect nonlinear transformation of the function $W(x,t)$, namely,
\be
v=\int^W_0t^n dt.
\ee
We shall see that the same solution was extremely easy to derive based on the similarity method here.

First we note that the form of $D=W^n$ implies a relation between the scaling exponents $d=nc$, which leads to the relation $\nu=n\mu$.
From $\nu=2\alpha-1$, one gets
\be
\alpha=-\mu=\frac{1}{n+2},~~n=0,1,2,\ldots
\ee

Next by integrating Eq.~(\ref{I-3}) once, one gets
\be
y^n(z)\frac{d}{dz}y(z)+\alpha z y(z)={\rm constant}
\label{I-3a}
\ee
The case with ${\rm constant} \neq 0$ is rather complicated, so here we shall follow \cite{Munier} and consider only the case with ${\rm constant}=0$. In this case, Eq.~(\ref{I-3a}) is easily integrated to give
\be
y(z)=\left(C-\frac{n\alpha}{2}z^2\right)^{\frac{1}{n}},~ C={\rm constant}.
\ee
From $W(x,t)=t^{\mu}y(z)$ and $\mu=-\alpha=-1/(n+2)$, we arrive easily at the solution given in \cite{Munier}
\be
W(x,t)=\left(Ct^{-\frac{n}{n+2}}-\frac{n}{2(n+2)}\frac{x^2}{t}\right)^{\frac{1}{n}}.
\ee

\be
%y(z)=\left[C-\frac{n}{2(n+2)}\frac{x
\ee

%%%%%%%%%%%%%%%%%%%%%
\section{Cases with $\mu\neq -\alpha$}

Now we come to the second type of scaling behaviour of the RDE, with $\mu\neq -\alpha$. For such RD system, the number of particles does not conserve.

The relevant equation is Eq.~(\ref{ODE})
\be
\rho y''+(\rho' +\alpha\,z)y'+\sigma-\mu\,y=0
\label{II}
\ee

%-----------
\subsection{$\rho=1$}

 In this case, Eq.\,(\ref{ODE}) becomes
 \be
 y^{\prime\prime}(z) +\alpha z y^\prime(z) -\mu y(z) + \sigma(y)=0.
 \label{II-1}
 \ee

 One can match this equation with any equation whose  solutions are exactly known. This will then determine the function $\sigma (y)$ and all the parameters.

 For instance, the differential equation  in $\S 2.1.2.29$ of \cite{PZ}
 \be
 y''+(az+b)y' +c[(a-c)z^2 +bz+1]y=0,
 \ee
 is known to have a particular solution
 \be
 y_0=e^{-\frac12 c z^2},
 \ee
for arbitrary constant $c$.  For our purpose, we must take $c>0$ to ensure normalizability of $W(x,t)$.

Comparing the two equations, one finds that for the choice
\be
a=\alpha, b=0,~c>0, ~~\sigma(y)=[c(\alpha-c)z^2+(\mu+c)]y_0(z),~~z\in (-\infty, \infty),
\ee
we obtain an exactly solvable RD system with
\bea
D(x,t)&=&t^{2\alpha-1},~~y(z)=e^{-\frac12 c z^2},~~x\in (-\infty, \infty),~\n\\
f(x,t)&=&t^{\mu-1}\left[c(\alpha-c)\left(\frac{x}{t^\alpha}\right)^2+(\mu+c)\right]e^{-\frac12 c \left(\frac{x}{t^\alpha}\right)^2},\label{II-1a}\\
W(x,t)&=& t^\mu e^{-\frac12 c \left(\frac{x}{t^\alpha}\right)^2}.\n
\eea
It can be checked that the continuity equation (\ref{EOC}) is
satisfied.

In Fig.3 we show the graphs of $D(x,t)$, $f(x,t)$ and $W(x,t)$ for
a set of parameters with three different times.

%-------------------
\subsection{$\rho=\beta\,z$}

The relevant equation is
\be
zy''(z)+\left(\frac{\alpha}{\beta} z+1\right)y' (z) +\frac{\sigma(y)-\mu y(z)}{\beta}=0.
\label{II-2}
\ee
This can be matched with the equation in
$\S 2.1.2.71$ of \cite{PZ}
\be
zy''+(az+b)y'+c[(a-c)z+b]y=0,
\ee
which has  a particular solution $y_0=e^{-cz}$ for $z\in [0,\infty)$. We shall take $c>0$ so that $y(z)\to 0$ as $x\to \infty$.
To match Eq.\,(\ref{II-2})  with this equation, one can take $a=\alpha/\beta, b=1$, and
\be
\sigma(y)=\left[c(\alpha-\beta\,c)z +\mu+\beta\, c\right]\,y_0.
\ee
Then
\bea
D(x,t)&=&\beta t^{\alpha-1} x,~~y(z)=e^{-\frac12 c z},~~ x\in [0, \infty),\n\\
f(x,t)&=&t^{\mu-1}\left[c(\alpha-\beta c)\,\frac{x}{t^\alpha}+(\mu+\beta c)\right]e^{-\frac12 c \frac{x}{t^\alpha}},\\
W(x,t)&=& t^\mu e^{-\frac12 c \frac{x}{t^\alpha}}.\n
\eea

%-------------------
\subsection{$\rho=-\frac12 \alpha z^2$}

Next we consider the choice $\rho=-\frac12 \alpha z^2$. This choice eliminates the first derivative term in $y(z)$. The RDE becomes
\be
y''-\frac{2}{\alpha}\left(\frac{\sigma-\mu y}{z^2}\right)=0.
\label{II-3}
\ee

%\subsubsection{g}

With the choice $\sigma(y)=\left[\frac12\alpha z^2 (g^2+g')+\mu\right]y(z)$, Eq.~(\ref{II-3}) reduces to
\be
y''-(g^2+g')y=0.
\ee
This equation admits a particular solution
\be
y_0(z)=e^{\int^z g(z) dz}.
\ee

We shall choose $g(z)$ such that $y(z)\to 0$ fast enough that the boundary terms go to zero.
For the above choice of $\sigma(y)$ and $y_0(z)$, we obtain an exactly solvable RD system with
\bea
D(x,t)&=&-\frac12\alpha \frac{x^2}{t},\n\\
f(x,t)&=&t^{\mu-1}\sigma(y),\label{II-2a}\\
W(x,t)&=& t^\mu \exp\left(\int^z g(z) dz\right).\n
\eea

Thus an exactly solvable RD system is obtained for every choice of $g(z)$ such that it is integrable. Below we shall present two simple examples.

%-----------
\subsubsection{$g=-1$}

In this case we have
\bea
y_0(z)&=&e^{-z},\n\\
f(x,t) &=&t^{\mu-1}\left[\frac{\alpha}{2}\left(\frac{x}{t^\alpha}\right)^2+\mu\right]e^{-\frac{x}{t^\alpha}},\\
W(x,t) &=&t^\mu e^{-\frac{x}{t^\alpha}},\n
\eea
for $t\geq 0$ and $x\in [0,\infty)$.

%---------
\subsubsection{$g=-z$}
In this case we have
\bea
y_0(z)&=&e^{-\frac12 z^2},\n\\
f(x,t) &=&t^{\mu-1}\left\{\frac{\alpha}{2}\left(\frac{x}{t^\alpha}\right)^2\left[\left(\frac{x}{t^\alpha}\right)^2-1\right]+\mu\right\}e^{-\frac12 \left(\frac{x}{t^\alpha}\right)^2},\\
W(x,t)&=&t^\mu e^{-\frac12 \left(\frac{x}{t^\alpha}\right)^2},\n
\eea
for $t\geq 0$ and $x\in (-\infty,\infty)$.

%-----------
\subsection{Nonlinear case: Generalized Fisher equation}

Let us consider the situation where the coefficients of Eq.~(\ref{II}) of $y''$ and $y'$ are proportional, say,
\be
\rho'+\alpha z=\gamma \rho.
\ee
The general solution of this equation is
\be
\rho(z)=\frac{\alpha}{\gamma}(z+\frac{1}{\gamma})+ \beta e^{\gamma z}, ~~\beta={\rm constant}.
\ee
Eq.~(\ref{II}) then reduces to
\be
y''+\gamma y' + \frac{1}{\rho}(\sigma-\mu y)=0.
\ee
An exactly solvable RDE can be obtained if one can match this equation with a known solvable ODE of the same form.
However, it can be easily checked that, to satisfy the continuity equation (\ref{EOCa}), the parameter $\alpha$ must be zero, i.e. $\alpha=0$, and thus $z=x,  \nu=-1$ and $D(x,t)=t^{-1}\beta e^{\gamma\,x}$.

As an example, let us consider the generalized Fisher equation
\be
y''+\gamma y'+ y(1-y^n)=0, ~~n>0,
\label{fisher}
\ee
which is an important equation in mathematical biology and nuclear physics \cite{Fisher,neutron,Murray}.
To match the RDE requires
\be
\sigma=\mu y+ \beta e^{\gamma\,x}  y\left(1-y^n\right).
\ee
The corresponding RDE is
\be
\frac{\partial W}{\partial t}=\frac{\partial}{\partial x}\left(\frac{\beta}{t}e^{\gamma\,x}\frac{\partial}{\partial x} W\right) + \frac{1}{t}\left[\mu\,W+\beta e^{\gamma\,x}W(1-t^{-n\mu}W^n)\right].
\ee
This equation is invariant under the scale transformation
\be
x={\bar x}, ~t=\epsilon^b\, {\bar t}, ~~W=\epsilon^{b\mu} \bar{W}.
\ee

It is known that particular solutions of Eq.~(\ref{fisher}) are possible if $\gamma=\pm (h_n+1/h_n)$, where $h_n\equiv \sqrt{n/2+1}$ \cite{K,W,R}.  The particular solutions are
\be
y^\pm (z)= \left(1+C e^{\pm(h_n-\frac{1}{h_n}) x}\right)^{-\frac{2}{n}},
\label{y}
\ee
where $C$ is some real constant.
We will take $C>0$ so that $W(x,t)$ is regular.
Also, the system defined by $\gamma<0$ is just the mirror image of that with $\gamma>0$, as Eq.~(\ref{y}) is invariant under $\gamma\to -\gamma$ and $x\to-x$.  So it suffices to consider just the case with $\gamma>0$.

In Fig.\,4 we show the graphs of $D(x,t), f(x,t)$ and $W(x,t)$ for a set of parameters with three different times and negative exponent $\mu<0$. $D(x,t)$ is independent of $\mu$, and $f(x,t)$ and $W(x,t)$ decrease as time increases.  For positive $\mu>0$, both $f(x,t)$ and $W(x,t)$ will increase as time increases. The traveling wave behaviour of the solution of the generalised Fisher equation is here transformed into a scaling wave behaviour.

\section{Summary}

We have considered  solvability of the reaction-diffusion
equation with both space- and time-dependent diffusion and reaction terms
by means of the similarity method. By introducing the
similarity variable, the reaction-diffusion equation is reduced to an
ordinary differential equation. It is interesting to realise that the reduced ordinary differential equations, namely, Eqs.\,(\ref{ODE}) and (\ref{ODE1}),  are quite simple in their functional forms. Particularly, Eq.\,(\ref{ODE1}) is integrable and its solution can be given in closed form. By matching these
two ordinary differential equations with known exactly solvable equations, one can obtain corresponding exactly solvable reaction-diffusion systems.   We have presented several representative examples of exactly solvable reaction-diffusion equation.

Of course, similarity solutions, just as the travelling wave solutions, are only one type of many possible symmetry solutions one can consider for the RDE.
Under different conditions,  the RDE may admit other symmetry solutions.
 Very recently, classification of exactly solvable RDE with gradient-dependent diffusivity (for $D$ a derivative of $W$, i.e., $D=D(W_x)$)  has been considered using the Lie symmetry approach \cite{CKK}.  It would be of interest to extend such consideration to more general diffusion and reaction terms.
% -----------------------------------------------------------------------------------------------------------------------------------------------------------------------------------------------

\acknowledgments

The work is supported in part by the Ministry of Science and Technology (MoST)
of the Republic of China under Grant NSC-102-2112-M-032-003-MY3.

%---------------------------------

%\bibliographystyle{plain}

%\newpage

%---------------  Figures ------

%---    Fig. 1 ---
\begin{figure}[ht] \centering
\includegraphics*[width=5cm,height=5cm]{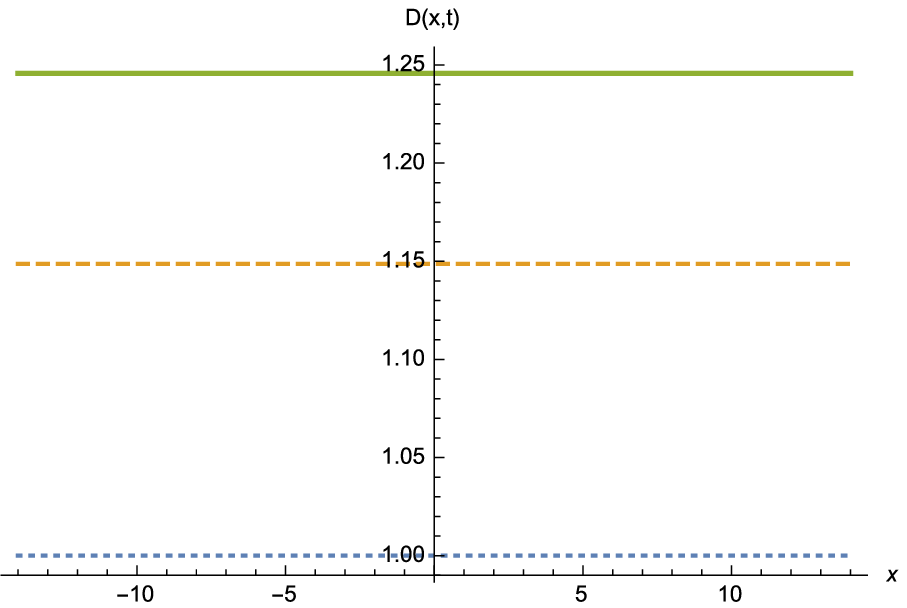}\hspace{1cm}
\includegraphics*[width=5cm,height=5cm]{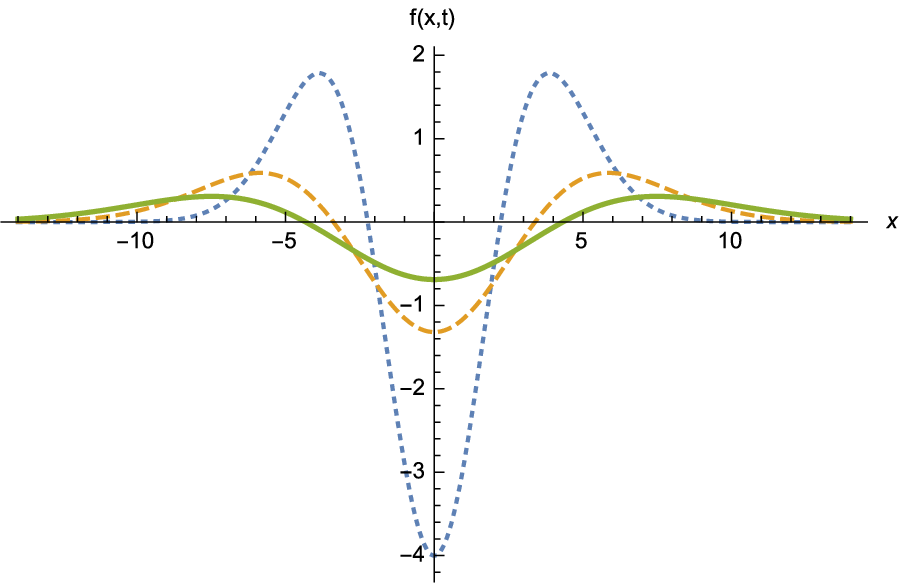}\hspace{1cm}
\includegraphics*[width=5cm,height=5cm]{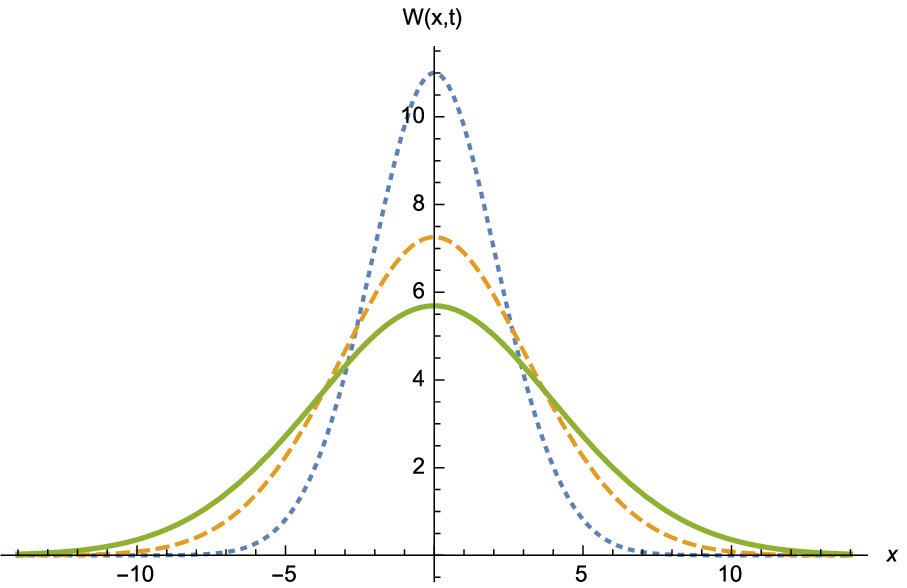}
\caption{Plot of $D(x,t), f(x,t)$ and $W(x,t)$ for $\alpha = 0.6, \gamma = 2, \eta = 0.1, C = $1 and time $t = 1.0$ (dotted), $t = 2.0$ (dashed), $t = 3.0$ (solid). }
\label{Fig1}
\end{figure}
%------------------

%---    Fig. 2 ---
\begin{figure}[ht] \centering
\includegraphics*[width=5cm,height=5cm]{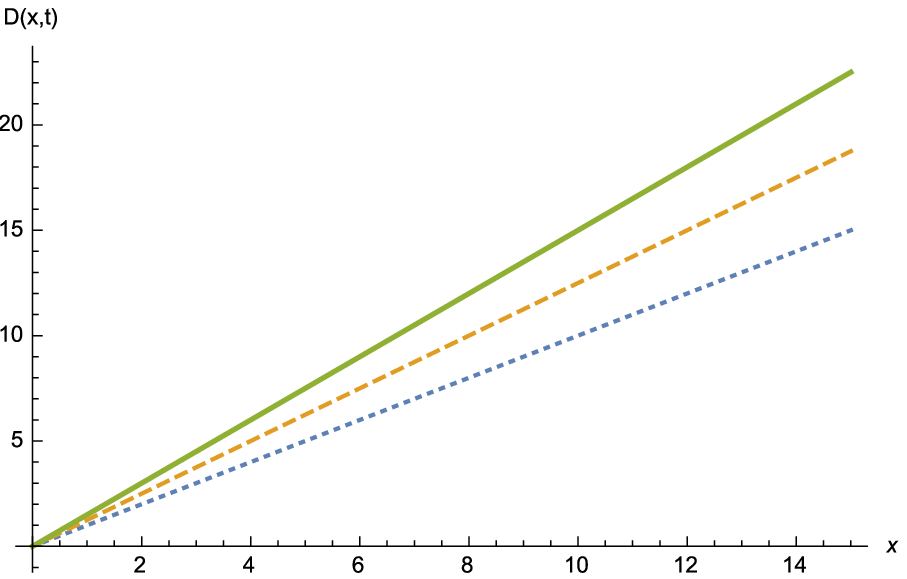}\hspace{1cm}
\includegraphics*[width=5cm,height=5cm]{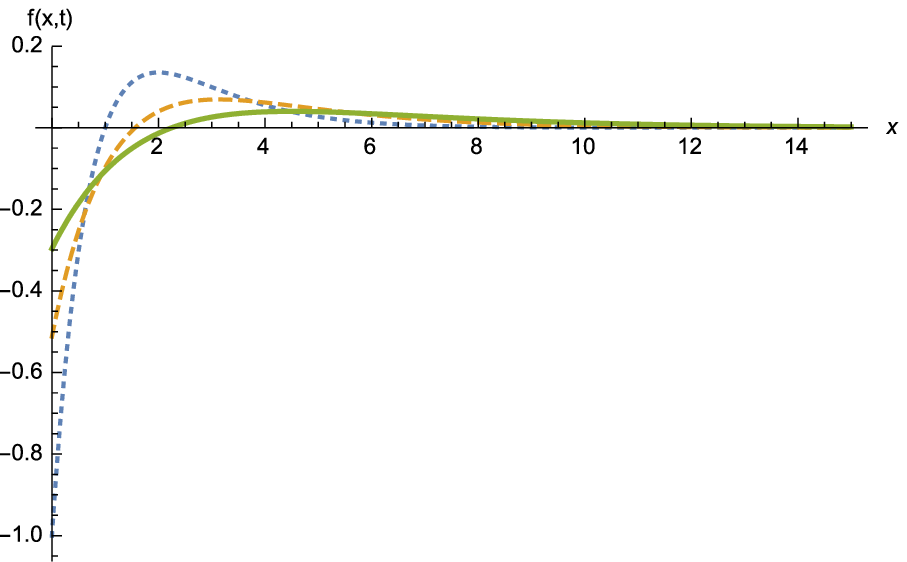}\hspace{1cm}
\includegraphics*[width=5cm,height=5cm]{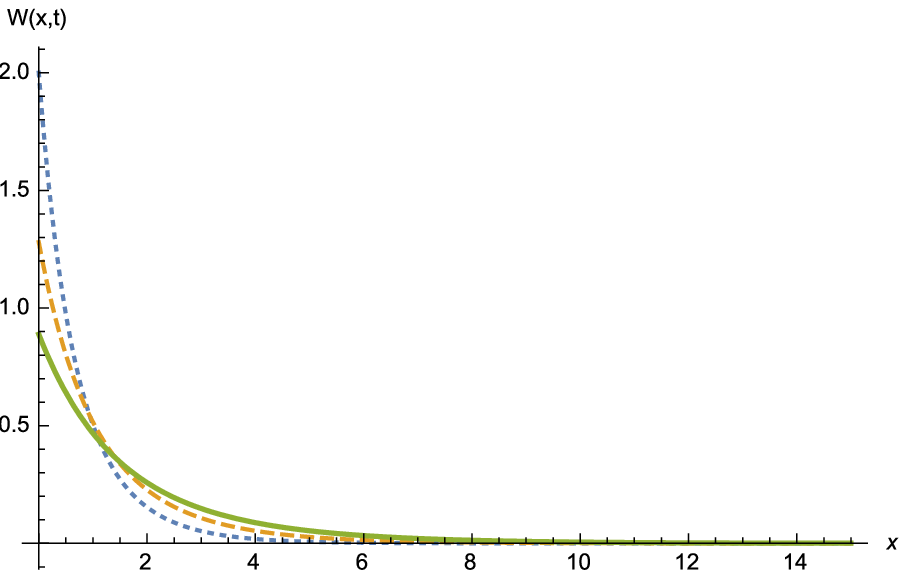}
\caption{Plot of $D(x,t), f(x,t)$ and $ W(x,t)$ for $\alpha = 2,
\eta = 1, C = 1$ and time $t = 1.0$ (dotted), $t = 1.25$ (dashed),
$t = 1.5$ (solid).} \label{Fig2}
\end{figure}

%------------------

%---    Fig. 3 ---
\begin{figure}[ht] \centering
\includegraphics*[width=5cm,height=5cm]{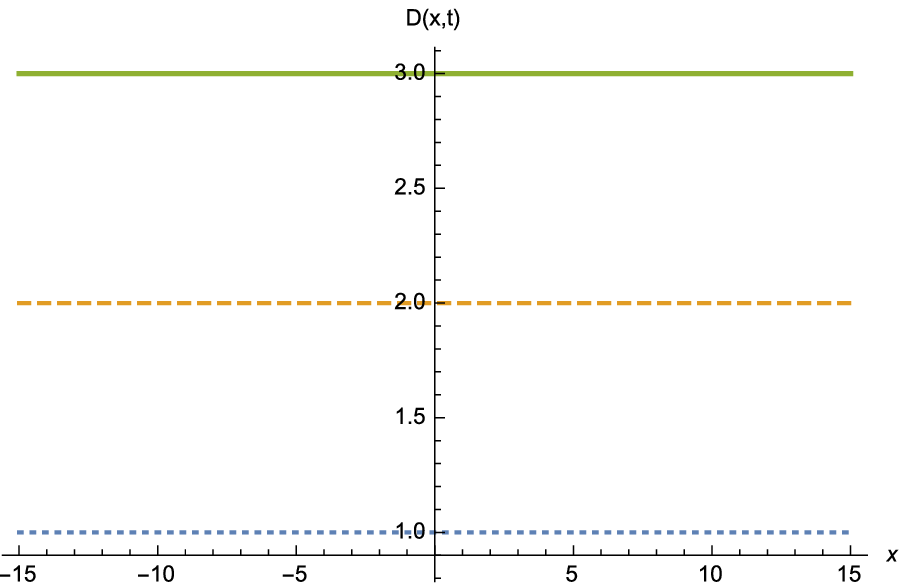}\hspace{1cm}
\includegraphics*[width=5cm,height=5cm]{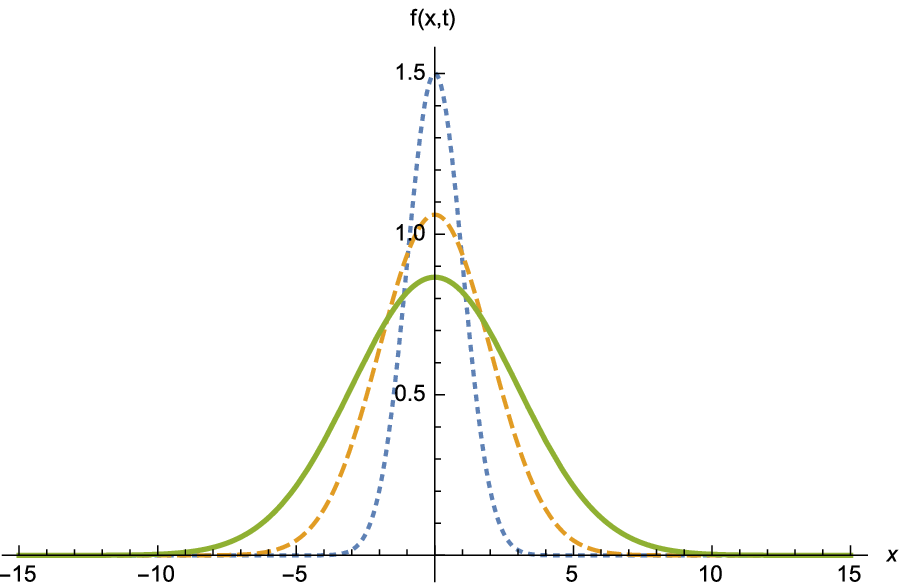}\hspace{1cm}
\includegraphics*[width=5cm,height=5cm]{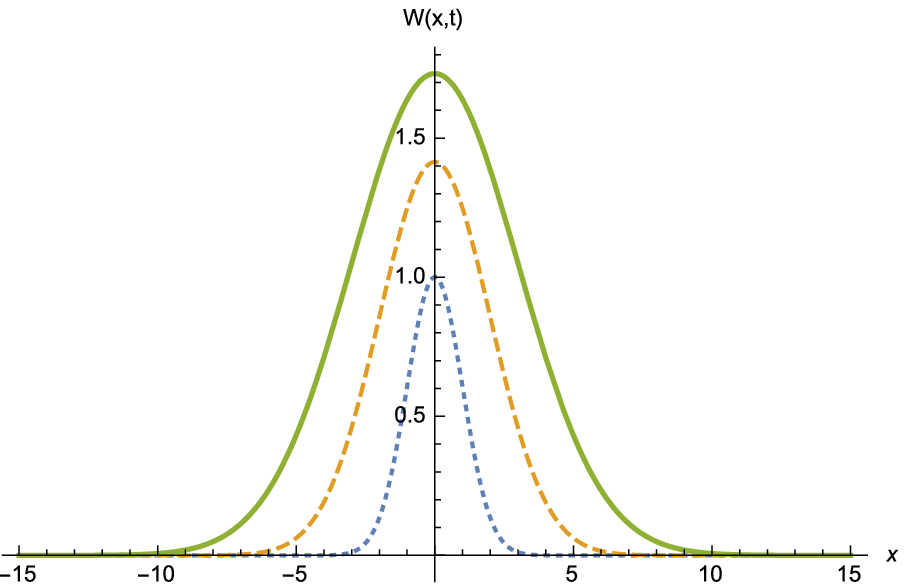}
\caption{Plot of $D(x,t), f(x,t)$ and $ W(x,t)$ for $\alpha = 1, \mu = 0.5, C = 1$  and time $t = 1.0$ (dotted), $t = 2.0$ (dashed), $t = 3.0$ (solid). }
\label{Fig3}
\end{figure}
%------------------

% ---  Fig. 4 -----
\begin{figure}[ht] \centering
\includegraphics*[width=5cm,height=5cm]{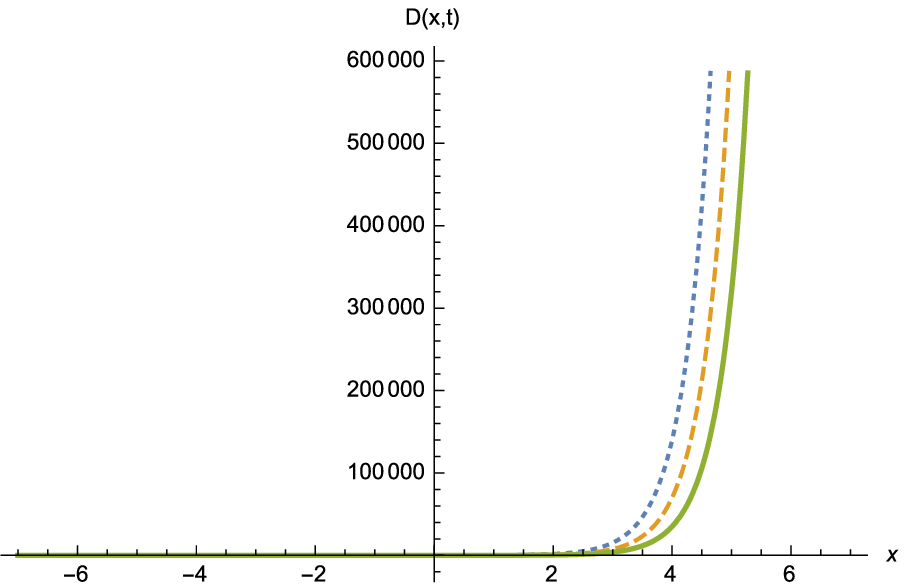}\hspace{1cm}
\includegraphics*[width=5cm,height=5cm]{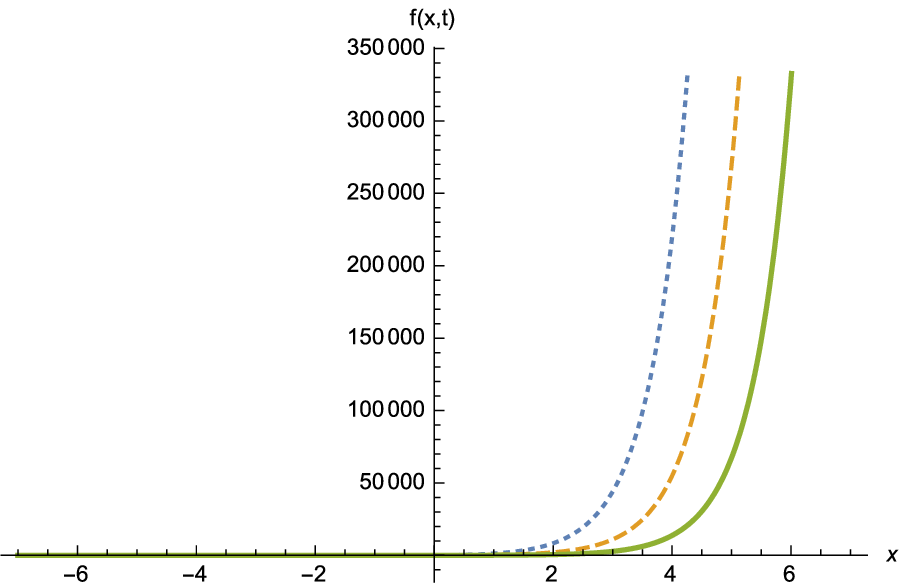}\hspace{1cm}
\includegraphics*[width=5cm,height=5cm]{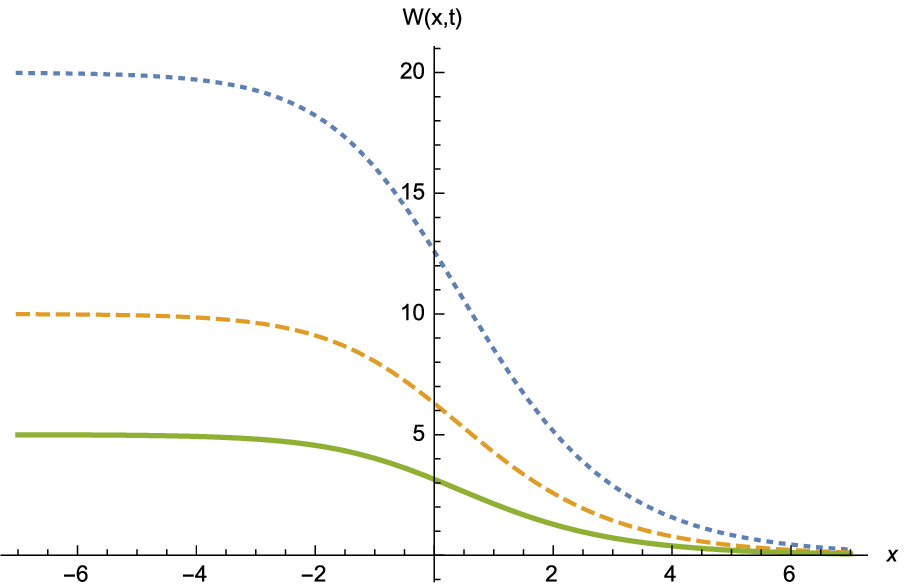}
\caption{Plot of $D(x,t), f(x,t)$ and  $W(x,t)$ for $n=3, \mu=-1, \beta=1, C=1$ and time $t=0.05$ (dotted), $0.1$ (dashed)  and $0.2$ (solid). }
\label{Fig4}
\end{figure}

%----------------------------


\begin{thebibliography}{99}

\bibitem{KPP}
A. Kolmogorov, I. Petrovsky and N. Piscunov, Bull. Moscow Univ.  A 1 (1937) 1.

%--- FPE -----
\bibitem{FPE}
H. Risken,  The Fokker-Planck Equation,  2nd. Ed., Springer-Verlag, Berlin, 1996.

%-----   Fisher eqn ----
\bibitem{Fisher}
R.A. Fisher, Ann. Eugenics 7 (1937) 353.
%The wave of advance of advantageous genes

%-- neutron evolution --
\bibitem{neutron}
J. Canosa, J. Math. Phys. 10 (1969) 1862.
% diffusion in nonlinear multiplicative media.

%---- Fluid dynamics ----
\bibitem{fluid}
A. C. Newell and J. A. Whitehead, J. Fluid Mech. 38 (1969) 279;\\
L. A. Segel, J. Fluid Mech. 38 (1969) 203.

%-- combustion and shock waves ---
\bibitem{shock}
Ya. B. Zeldovich and D. A. Frank-Kamenetsky, Acta Physicochim. 9 (1938) 341;\\
J. Smoller, Shock Waves and Reaction Diffusion Equations, Springer (1994).

% -- chemical kinetics --
\bibitem{Arnold}
R. Arnold, K. Showalter and J.J. Tyson, J. Chem. Educ. 64 (1987) 740
% Propagation of chemical reactions in space


%---  pattern formation ---
\bibitem{Turing}
A.M.  Turing,  Phil. Trans. Roy. Soc. Lond. B237 (1952) 37.
% The chemical basis of morphogenesis

%-- nerve systems ---
\bibitem{nerve}
A. L. Hodgkin and A. F. Huxley, J. Physiol. (Lond.) 117 (1952) 500;\\
R. FitzHugh, Biophys. J. 1 (1961) 445;\\
J. Nagumo, S. Arimoto and S.  Yoshizawa, Proc. Inst. Radio Eng. 50 (1962) 2061.

%---   Biology ---
\bibitem{Murray}
J.D. Murray, Mathematical Biology, 2nd Ed., Springer-Verlag, Berlin, 1993.

%---  travelling waves ----
\bibitem{GK}
B. H. Gilding and R. Kersner, Travelling Waves in Nonlinear Diffusion Convection Reaction, Birkhäuser, Springer, 2004.

%----    Similarity methods -----
\bibitem{BC}
G. W. Bluman and J. D. Cole, Similarity Methods for
Differential Equations, Springer-Verlag, New York, 1974.

% -----    our previous papers -------
\bibitem{Ho1}
W.-T. Lin and C.-L. Ho,
%Similarity solutions of Fokker-Planck Equations with time-dependent coefficients,
Ann. Phys. 327 (2012) 386.
% Issue 2, February 2012, Pages 386-397
%Tamkang preprint, Jun 2011 (to appear in Ann. Phys).

\bibitem{Ho2}
 C.-L. Ho, %Similarity solutions of Fokker-Planck equation with moving boundaries,
J. Math. Phys. 54  (2013) 041501.

\bibitem{Ho3}
 C.-L. Ho and R. Sasaki,
%Extensions of a class of similarity solutions of Fokker-Planck equation with time-dependent coefficients and fixed/moving boundaries,
J. Math. Phys. 55 (2014) 113301.

%----  nonlin. Diffusion eqn ----
\bibitem{Munier}
A. Munier, J.R. Burgan, J. Gutierrez, E. Fijalkow and M.R. Feix, , SIAM J. Appl. Math 40 (1981) 191.

%---    Handbook ----

\bibitem{PZ}
A.D. Polyanin and V. F. Zaitsev, Handbook of Exact Solutions for Ordinary Differential Equations, 2nd ed., Chapman $\&$ Hall/CRC, London, 2003.

%------   Fisher eqn ----

%\bibitem{Fisher}
%R.A. Fisher, Ann. Eugenics 7 (1937) 353.
%The wave of advance of advantageous genes

\bibitem{K}
P. Kaliappan, Physica 11D (1984) 368.

\bibitem{W}
X.Y. Wang, Phys. Lett. A131 (1988) 277.

\bibitem{R}
H.C. Rosu and O. Cornejo-P\'erez, Phys. Rev. E 71 (2005)  046607.

%---   Lie approach --
\bibitem{CKK}
R. Cherniha, H.R. King and S. Kovalenko, Lie symmetry properties of nonlinear reaction-diffusion equations with gradient-dependent diffusivity,
arXiv: 1507:01893 [math-ph].

\end{thebibliography}
\end{document}